\documentclass[floatfix,%
preprint,
preprintnumbers,
 amsmath,amssymb,
 aps,
]{revtex4-2}
\usepackage{siunitx}
\usepackage{graphicx}% Include figure files
\usepackage{dcolumn}% Align table columns on decimal point
\usepackage{bm}% bold math
\usepackage{xcolor}
\usepackage{amsmath}
\usepackage{booktabs}
\usepackage{hyperref}% add hypertext capabilities
\usepackage{tikz}
\usetikzlibrary{shapes.geometric, positioning}
\usepackage{placeins}

\begin{document}

%\preprint{APS/123-QED}

\title{Multiphysics Analysis of Cryogenically Cooled Photocathode in a CW SRF Injector cavity}% 

\thanks{Work performed in the framework of R$\&$D for future accelerator
operation modes at the European XFEL and financed by the
European XFEL GmbH.}%

\author{Dmitry Bazyl}
  \email{dmitry.bazyl@desy.de}
  \author{Klaus Floettmann}%
  \author{Elmar Vogel}%
\author{Igor Zagorodnov}%

%\author{}%

\affiliation{%
Deutsches Elektronen-Synchrotron DESY, Notkestrasse 85, 22607 Hamburg, Germany
}%

\date{\today}% 

\begin{abstract} 

The paper evaluates the thermal regime of a cryogenically cooled copper photocathode integrated into a continuous-wave superconducting radio-frequency injector cavity with direct thermal contact. Such a photoinjector layout is being developed at DESY and has recently demonstrated a record-high 50 MV/m axial electric field in radio-frequency tests, marking an important milestone. To address the thermal effect of the picosecond excitation laser, we first develop a two-temperature model to describe the temperature of the emitting surface at cryogenic temperatures and solve it numerically. Subsequently, we present a one-temperature model of the bulk photocathode coupled with an electromagnetic model of the injector cavity. For the current injector design, we predict a negligible impact of the laser on the intrinsic quality factor of the cavity, identifying instead the cryogenic stability of the copper cathode as the primary operational limit. To overcome cooling challenges, we propose an improved configuration of the cathode plug. For the proposed geometry, the multiphysics analysis confirms stable performance at a nominal 2 W laser power, sufficient for 100 pC beams at 1 MHz under optimistic quantum efficiency assumptions. Operation at higher laser loads will benefit from further dedicated cryogenic analysis.

\end{abstract}

\maketitle

\tableofcontents

\section{\label{sec:intr}INTRODUCTION}
A superconducting radio-frequency (SRF) photoinjector is being developed at DESY aiming to enable high-duty-cycle / continuous wave (CW) operation in a future upgrade of the European X-ray Free-Electron Laser~\cite{Decking2020,BRINKMANN201420}. The core feature of the DESY CW SRF injector design is a photocathode insertion method~\cite{Vogel:2018piq}. The cathode is thread-mounted to the backside of the SRF injector cavity made of superconducting niobium. This design has experimentally demonstrated an axial electric field up to 50 MV/m in radio-frequency vertical tests (VT). The result supports the fundamental feasibility of generating 100 pC electron bunch charge with a transverse slice emittance below \SI{0.2}{\micro\meter}~\cite{bazyl2021cw}. Such photocathode insertion method omits complexities related to load-lock systems and overall such an injector type serves as a promising candidate to fulfill many demanding scientific cases beyond electron sources for CW linacs driving FELs such as CW ultrafast electron diffraction (UED)~\cite{RevModPhys.94.045004} and CW THz radiation~\cite{Tonouchi2007} sources. A unique feature of this photoinjector is the direct 2 Kelvin cooling interface of the photocathode via superfluid helium. 

A limited  number of studies address the characteristics of copper photocathodes at cryogenic temperatures. Most available data concern photoemissive properties of copper such as quantum efficiency (QE) and mean transverse energy (MTE). One study showed that the QE of polycrystalline Cu decreases by a factor of four when the substrate temperature is lowered from 400~K to 85~K~\cite{Harris2013}. Tests at the bERLinPro injector demonstrated a QE of $10^{-5}$ at 257~nm with an Cu insert at 80~K~\cite{Neumann2018}. The record low-emittance benchmark for Cu(100) with 5~meV MTE at 35~K was reported in~\cite{Karkare2020}. 

In the context of photoinjector systems operating in the cryogenic range, thermal stability of the SRF cryogenic assembly of the injector cavity and the photocathode is a necessary prerequisite for stable operation. Previous research on the SRF injector thermal management has mostly focused on systems, incorporating load-lock mechanisms for photocathode insertion. For such systems, thermal simulations and measurements have been performed to study the heat generated on a retractable cathode stalk, which acts as the primary heat leak and requires intermediate cooling stages (e.g., 77K intercepts or active gas cooling) as reported in ~\cite{Kuehn2019TUP100}. For the same injector type with a load-lock system, the finite element method (FEM) has been implemented to carry out thermal studies \cite{Kostin2015Breakdown}. Closer to our direct cathode mount configuration, Schultheiss et al. \cite{Schultheiss2000Thermal} studied the thermal behavior of an SRF injector cavity operated at 4.2 K, made of residual-resistance ratio (RRR) 300 niobium, where the photocathode is part of the cavity's backside. The injector cavity under RF load and 1 W laser heating was analyzed, and it was shown that  the injector cavity temperatures remained below the superconducting critical temperature. The application of FEMs for thermal modeling at cryogenic temperatures is robust with results often matching experimentally obtained data~\cite{Barbanotti2008_TTFcooldown, Awida2020_ThermalQuench}.
 
\begin{figure}[ht]
\includegraphics[width=1.0\linewidth]{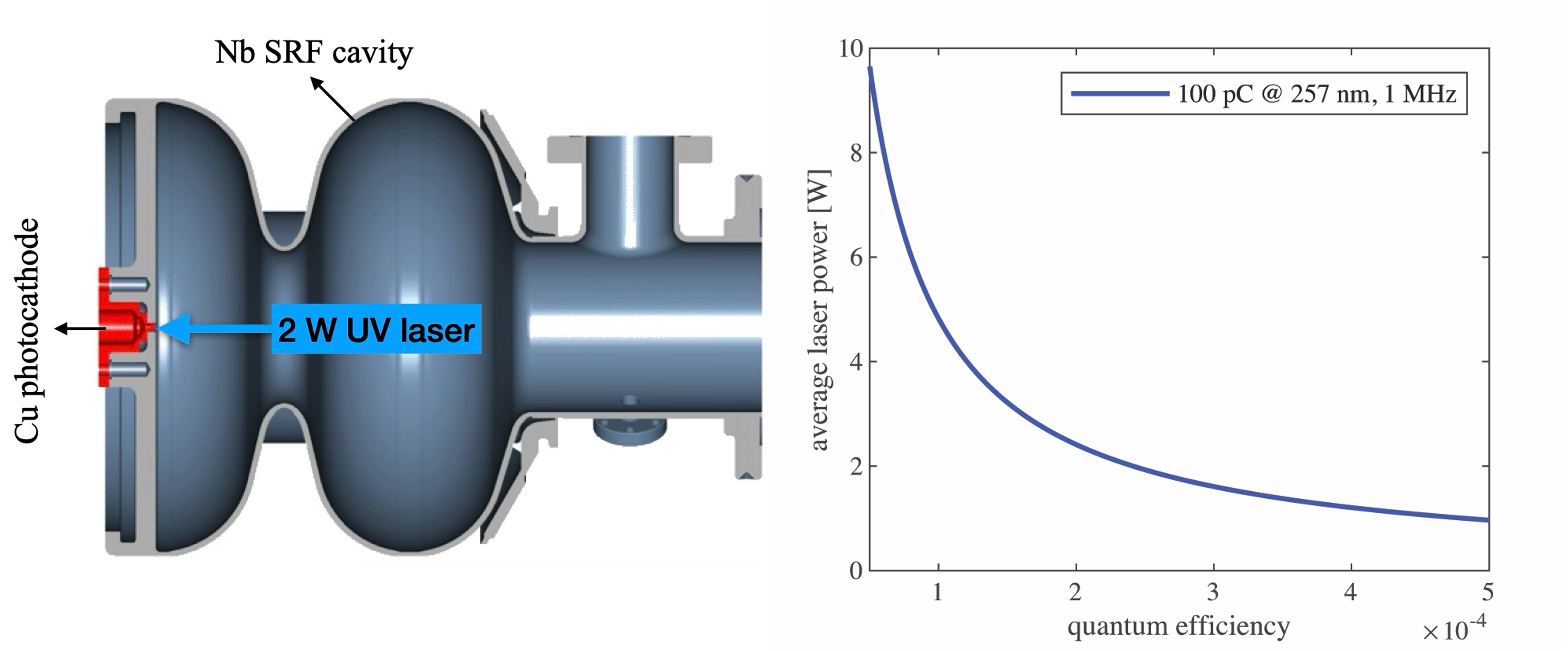}
\caption{The image on the left shows the three-dimensional layout of the CW SRF injector cavity with an integrated copper photocathode. The plot on the right presents the average laser power required for a Cu photocathode as a function of QE, assuming a $1 \mathrm{MHz}$ repetition rate and $100 \mathrm{pC}$ bunch charge (257 nm excitation laser).}
\label{fig:gun_model}
\end{figure}

 The CW SRF photoinjector with backside thread-mounting of the cathode (see Fig. \ref{fig:gun_model}) assumes that most of the heat transfer from the Cu cathode to the superconducting Nb cavity occurs indirectly via an indium sealing. The dissipated RF power in the SRF cavity $P_{\text{d}}$ is directly proportional to the surface resistance of the conducting material. The surface resistance $R_s(T)$ of superconducting niobium is a combination of the residual resistance and the temperature-dependent (Bardeen–Cooper–Schrieffer) BCS resistance~\cite{Padamsee2008RFAccelerators}. The dissipated power is related to the total surface resistance and the quality factor as follows \cite{wangler2008rf}

\begin{equation}
 P_{\text{d}}(T) = \frac{1}{2} \int_S R_s(T) |\mathbf{H}_t|^2 dS, \qquad Q_0 = \frac{\omega U}{P_{\text{d}}}.
    \label{eq:pd_q0}
\end{equation}
where $|\mathbf{H}_t|$ is the magnitude of the tangential magnetic field at the surface $S$, $\omega$ is the angular resonant frequency, and $U$ is the energy stored in the cavity.

For FEL applications, a Cu photocathode requires up to 5 W of laser power to generate 100 pC electron bunches with a repetition rate up to 1 MHz under the optimistic assumption of a QE range for Cu of $(1-2.5)\times10^{-4}$ at 257~nm  as shown in the right plot of Fig. \ref{fig:gun_model}. For~comparison, the RF generated power loss distributed over the backside of the SRF injector cavity is around 1 W for 50 MV/m axial field. Therefore, a substantial laser power deposited into the cryogenic system must be addressed to confirm the thermal stability as well as the absence of a quality factor degradation due to potentially increased RF losses.

The problem of the thermal stability of a CW SRF injector cavity with a Cu cathode integrated via a thread-mounted connection is unique and therefore has not been previously addressed in the literature. To our knowledge, no prior work has solved a two-temperature model for copper initialized in the cryogenic regime under UV excitation. To address these questions, we have developed the required models and used  \textsc{Comsol Multiphysics\textsuperscript{\textregistered}}~v6.3 \cite{comsol} to analyze the complex thermal problem of the cold cathode integrated into the SRF injector cavity. 

The analysis reveals a potential cryogenic instability in the baseline cathode plug geometry\cite{Sekutowicz2025}, leading us to propose and evaluate a modified design for enhanced thermal stability at a nominal incident laser power of 2 W.

We start in Sec.~\ref{sec:1dttm} by developing a two-temperature model, to describe the temperature of the emitting surface at cryogenic temperatures and solve it numerically, to characterize the surface temperature evolution of the cryogenically cooled Cu photocathode during picosecond ultraviolet (UV) laser irradiation and to obtain the time-dependent heat flux for the macroscopic model. In Sec.~\ref{sec:2dSTM} we analyze the two-dimensional transient thermal behavior of the bulk cathode and assess the validity of steady-state thermal analysis in regions distant from the effective emission surface (the Cu-Nb contact interface). Finally, in  Sec.~\ref{sec:coupled} we evaluate a coupled three-dimensional thermal-electromagnetic model where the influence of heat induced by the photocathode laser on the dissipated RF power is studied.

\section{\label{sec:1dttm}One-dimensional Transient thermal analysis\\ of a photoemissive surface}

In order to capture the ultrafast thermal dynamics in a copper photocathode at cryogenic temperatures during and after laser irradiation, we employ a one‑dimensional two‑temperature model (TTM). It resolves the electron temperature $T_e$ and lattice temperature $T_l$ as a function of depth~$x$ from the irradiated surface~\cite{Anisimov1974JETP} and reads 
\begin{align}
C_e(T_e) \frac{\partial T_e}{\partial t} &= \frac{\partial}{\partial x} 
\left[ K_e(T_e, T_l) \frac{\partial T_e}{\partial x} \right]
- G(T_e, T_l)(T_e - T_l) + S_{\text{laser}}(x,t)
\label{eq:TTM_Te}
\end{align}
\begin{equation}
C_l(T_l) \frac{\partial T_l}{\partial t} = \frac{\partial}{\partial x}\left[ K_l(T_l) \frac{\partial T_l}{\partial x}\right] + G(T_e,T_l)(T_e-T_l)
\label{eq:TTM_Tl}
\end{equation}
where $C_e$ and $C_l$ are the volumetric heat capacities of the electron and lattice subsystems, respectively; $K_e$ and $K_l$ are the corresponding thermal conductivities; $G$ is the electron-phonon coupling factor that quantifies the volumetric rate of energy exchange between the two subsystems; and $S_{\text{laser}}$ is the volumetric laser heat source term.

The domain spans $x\in[0,L_x]$ with $L_x=50\,\mu\mathrm{m}$. The relatively large simulation domain (compared to typical conditions for TTM models at room temperatures) is motivated by the exceptionally low heat capacity and high thermal conductivity of cryogenic copper, which leads to a rapid heat transport. Thus, we ensure that the ultrafast near-surface electron and lattice temperature dynamics, which govern the net electron-phonon energy exchange, are not prematurely influenced by the thermal conditions imposed at the rear boundary. The initial conditions are given by $T_e(x,0)=T_l(x,0)=T_0$ ($T_0=2 \mathrm{K}$). A zero‑flux boundary condition is applied at $x=0$, while Dirichlet boundary condition $T_e=T_l=T_0$ is imposed at $x=L_x$.

A two-temperature model for a copper photocathode under ultrashort laser pulse excitation has previously been considered for room-temperature copper in \cite{Maxson2017}. To validate our numerical model, we successfully reproduced selected results.

In the following subsections, we extend this model to our case of interest at cryogenic temperatures. We have conducted a review of the available literature to combine theoretical and experimental results in order to obtain the required coefficients at cryogenic temperatures for the 1D-TTM. The obtained equations are solved numerically for a representative laser-source configuration defined by the use-case scenario. The results are then applied in Sec.~\ref{sec:2dSTM} in the one-temperature modeling of the injector.

\subsection{\label{sec:mat_prop}Properties of copper at cryogenic temperatures}
In the following subsections, we provide analytical definitions of the material properties of copper required to solve the 1D-TTM in the cryogenic temperature range. The lattice thermal conductivity $K_l(T_l)$ is omitted due to its negligible contribution for high-purity copper across the entire temperature range and the fact that $K_e \gg K_l$ \cite{PhysRevB.100.144306}.

\paragraph{Electron–Phonon Coupling Factor $G$.}
The electron-phonon coupling factor $G(T_e,T_l)$ quantifies the volumetric rate of energy exchange between the electron and lattice subsystems. While $G$ is often treated as a constant in room-temperature TTM, its strong temperature dependence requires an accurate description of the thermal dynamics which starts from the cryogenic regime. At very low temperatures (significantly lower than the Debye temperature, $T \ll \Theta_D$), the energy transfer rate per unit volume, $P_{ep}/V$, between electrons at $T_e$ and a lattice at $T_l$ in clean, three-dimensional metals is theoretically predicted to follow a power law \cite{Kaganov1957, Allen1987prl, Wellstood1994prb}. Specifically, for energy transfer limited by electron interactions with 3D acoustic phonons, this rate is given by:
\begin{equation}
    P_{ep}/V = \Sigma_0 (T_e^5 - T_l^5),
    \label{eq:Pep_lowT}
\end{equation}
where $\Sigma_0$ is a material-specific coupling coefficient. The $T^5$ dependence arises from the temperature dependence of both the available phonon states and the electron phase space for scattering \cite{RevModPhys.78.217}.
The electron-phonon coupling factor $G(T_e,T_l)$ is defined such that $P_{ep}/V = G(T_e,T_l)(T_e-T_l)$. Thus, from Eq.~(\ref{eq:Pep_lowT}), the low-temperature form of $G$ is:
\begin{equation}
    G_{\mathrm{low}}(T_e,T_l) = \Sigma_0 \sum_{k=0}^4 T_e^{4-k} T_l^{k}.
    \label{eq:Glow_def}
\end{equation}
For copper, the material coefficient $\Sigma_0 = \SI{2.1e9}{\watt\per\meter\cubed\per\kelvin\tothe{5}}$ is adopted. This value is supported by direct experimental measurements on copper thin films that exhibit 3D phonon behavior at sub-Kelvin temperatures \cite{Wang2019apl}, and is also consistent with earlier work on similar metallic systems \cite{Wellstood1994prb}. For instance, if $T_e=T_l=T$, then $G_{\mathrm{low}}(T,T) = 5 \Sigma_0 T^4$. At $T=\SI{2}{K}$, this yields $G_{\mathrm{low}} \approx \SI{1.68e11}{\watt\per\meter\cubed\per\kelvin}$, and at $T=\SI{27}{K}$, $G_{\mathrm{low}} \approx \SI{5.58e15}{\watt\per\meter\cubed\per\kelvin}$. 

For temperatures approaching the Debye temperature (343 K for Cu), $G(T_e,T_l)$ tends towards a less sharply temperature-dependent reference value, $G_{\mathrm{RT}}$. For copper near \SI{300}{K}, a value of $G_{\mathrm{RT}} = \SI{9.0e16}{\watt\per\meter\cubed\per\kelvin}$ is adopted, consistent with pump-probe experimental results \cite{10.1063/1.5035368, Hohlfeld2001chemphys} and the baseline $G_0$ derived from Density Functional Theory (DFT) calculations before strong electron temperature effects dominate \cite{Lin2008prb}. While some models predict a decrease in G for electron temperatures far exceeding the Debye temperature \cite{Lin2008prb}, treating $G_{RT}$ as a constant saturation value is a valid approximation for the temperature regimes relevant to this work.

\begin{figure}[h]
\includegraphics[width=0.5\linewidth]{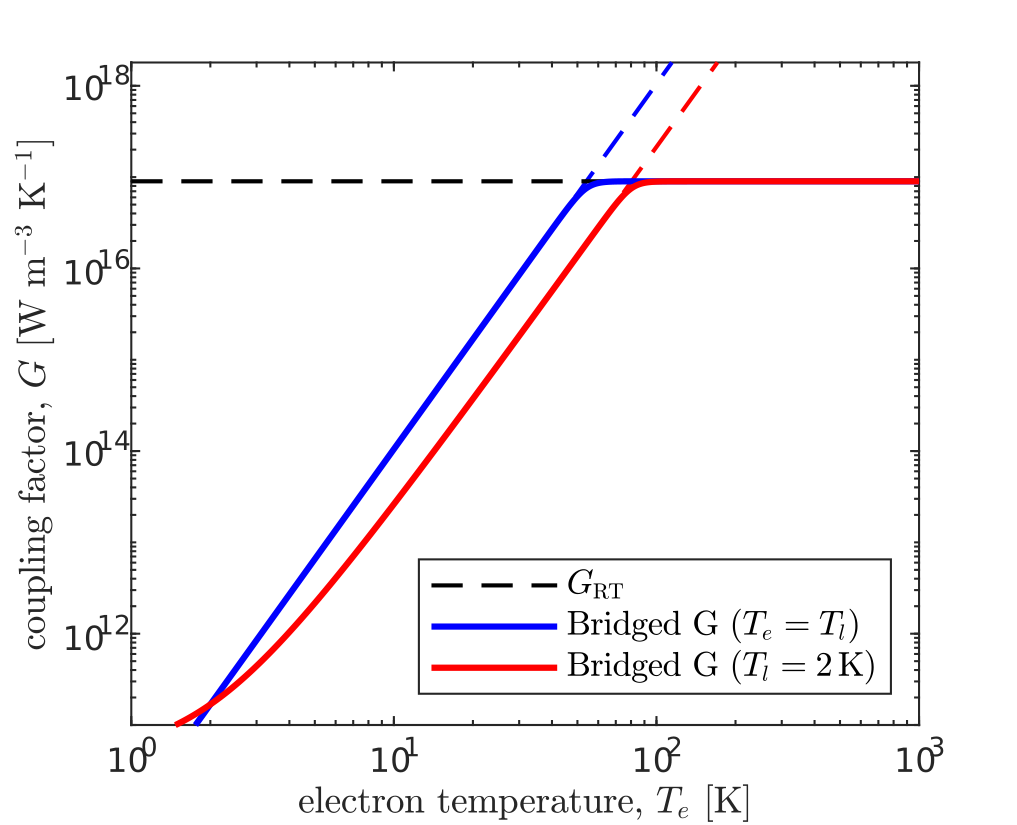}
\caption{The electron-phonon coupling factor, $G$, as a function of electron temperature. It illustrates the behavior of the full bridged model under two limiting conditions: thermal equilibrium ($T_e = T_l$) and strong non-equilibrium ($T_l=2\,$K, varying $T_e$). The asymptotic low-temperature power-law ($G_{\text{low}}$) and high-temperature saturation ($G_{\text{RT}}$) models are shown for reference.}
\label{fig:bridge}
\end{figure}

To provide a smooth transition across the entire temperature range from the low-temperature power-law behavior to the higher-temperature reference value, two limiting forms are combined using a generalized mean:
\begin{equation}
    G(T_e,T_l) = \left[ \left(G_{\mathrm{low}}(T_e,T_l)\right)^{-n_b} + \left(G_{\mathrm{RT}}\right)^{-n_b} \right]^{-1/n_b},
    \label{eq:G_coupling_bridged_simplified} % 
\end{equation}
with a bridging exponent $n_b=4$ chosen to provide a smooth and monotonic transition between the two asymptotic regimes as illustrated in Fig. \ref{fig:bridge}. The formulation dictates that $G_{\mathrm{low}}$ dominates at temperatures below the natural crossover point (approximately \SI{54}{\kelvin} for $T_e=T_l$), and $G_{\mathrm{RT}}$ primarily governs the coupling at higher temperatures.

\paragraph{Electron heat capacity $C_e$.}

The volumetric heat capacity of the electron gas, $C_e(T_e)$, is commonly approximated at low temperatures by its linear term:
\begin{equation}
    C_e(T_e) = \gamma_e T_e,
    \label{eq:Ce_linear}
\end{equation}
where $\gamma_e$ is the electronic specific heat coefficient. For copper, $\gamma_e \approx \SI{96.6}{\joule\per\metre\cubed\per\kelvin\squared}$. The linear approximation, provides good accuracy for relatively low electron temperatures which are relevant to this work. Worth noting, when electron temperatures $T_e$ become more elevated and approach values in the order of several thousand Kelvin it becomes advisable to introduce higher-order terms from the Sommerfeld expansion to Eq. \ref{eq:Ce_linear} for improved accuracy\cite{Ashcroft1976,Kittel2005}. 

\paragraph{Lattice heat capacity $C_l$.}
The volumetric heat capacity of the lattice is determined using the Debye model \cite{Kittel2005}:
\begin{equation}
C_l(T_l)=9N_a k_B\left(\frac{T_l}{\Theta_D}\right)^{3} 
\times \int_0^{\Theta_D/T_l} \frac{\xi^4 e^\xi}{(e^\xi-1)^2}\,\mathrm{d}\xi,
\end{equation}
where $N_a = \SI{8.4912e28}{\per\metre\cubed}$ is the number density of atoms for copper, $k_B$ is the Boltzmann constant, and $\Theta_D = \SI{343}{\kelvin}$ is the Debye temperature for copper.

\paragraph{Electrical Resistivity $\rho_{el}$.}
The total electrical resistivity $\rho_{el}(T_l)$ is determined by the Matthiessen's rule, summing the temperature-independent residual resistivity $\rho_0$ (due to impurities and defects) and the temperature-dependent phonon contribution $\rho_{ph}(T_l)$ \cite{Ziman1960}:
\begin{equation}
    \rho_{el}(T_l) = \rho_0 + \rho_{ph}(T_l).
\end{equation}
The residual resistivity $\rho_0 = \SI{5.667e-11}{\ohm\metre}$ is set corresponding to a RRR of 300, using the reference resistivity of copper at room temperature $\rho_{el}(\SI{300}{\kelvin}) = \SI{1.7e-8}{\ohm\metre}$. 

The phonon contribution is described by the Bloch-Grüneisen formula:
\begin{equation}
    \rho_{ph}(T_l) = A_{\mathrm{BG}} \left(\frac{T_l}{\Theta_D}\right)^{5}  \times \int_0^{\Theta_D/T_l} \frac{\xi^5 e^\xi}{(e^\xi-1)^2} \, \mathrm{d}\xi.\end{equation}
Prefactor $A_{\mathrm{BG}}$ is determined to ensure that $\rho_0 + \rho_{ph}(\SI{300}{\kelvin})$ yields the room temperature value. 

\paragraph{Electron thermal conductivity $K_e$.}
The electron thermal conductivity $K_e(T_e,T_l)$ is derived from the Wiedemann-Franz law \cite{Ashcroft1976}, relating it to the electrical resistivity:
\begin{equation}
K_e(T_e,T_l)= \frac{L_0 T_e}{\rho_{el}(T_l)},
\end{equation}
where $L_0 = \SI{2.44e-8}{\watt\ohm\per\kelvin\squared}$ is the Lorenz number. The lattice temperature dependence enters through the electrical resistivity $\rho_{el}(T_l)$.

%, is implemented in our work as a cubic-spline interpolation function, derived from experimental data and theoretical calculations for copper , as shown in the right plot of Fig.~\ref{fig:bridge}. This  reproduces the characteristic $T_l^2$ behavior below $10 \, \text{K}$, a peak value of approximately $15 \, \text{W} \cdot \text{m}^{-1} \cdot \text{K}^{-1}$ (mid-range of $10\text{--}20 \, \text{W} \cdot \text{m}^{-1} \cdot \text{K}^{-1}$) near $34 \, \text{K}$ ($T_l \approx \Theta_D/10$), and the subsequent monotonic decrease due to Umklapp scattering to approximately $7\text{--}8 \, \text{W} \cdot \text{m}^{-1} \cdot \text{K}^{-1}$ at $300 \, \text{K}$.

\subsection{Laser source term}
The volumetric power density $S_{\mathrm{laser}}(x,t)$ deposited by the laser into the electron system is described by a Gaussian temporal profile with an exponential decay in depth according to the Beer-Lambert law:

\begin{equation}
\begin{split}
S_{\mathrm{laser}}(x,t) ={}& \frac{F_{\mathrm{abs}}}{d_p} \frac{1}{\sigma_t\sqrt{2\pi}} \exp\left[-\frac{(t-t_c)^2}{2\sigma_t^2}\right] \exp\left(-\frac{x}{d_p}\right)
\end{split},
 \label{eq:source_term}
\end{equation}
where $F_{\mathrm{abs}}$ is the absorbed laser fluence, $d_p$ is the effective penetration depth, $\sigma_t$ is the rms pulse duration, and $t_c$ is the time corresponding to the pulse peak. The variables $x$ and $t$ represent the depth into the material and the time, respectively.

\subsection{Numerical solution of 1D-TTM}
The simulation models the interaction of a single laser pulse with a copper surface in 1D. Equations~(\ref{eq:TTM_Te}) and~(\ref{eq:TTM_Tl}) were defined as custom  partial differential equations and solved in \textsc{Comsol Multiphysics\textsuperscript{\textregistered}}~v6.3 using the adaptive backward differentiation formula (BDF) solver. A non‑uniform mesh resolves the optical penetration depth near $x=0$ and coarsens toward $x=L_x=\SI{50}{\mu m}$. 

\begin{figure}[htb]
\includegraphics[width=0.5\linewidth]{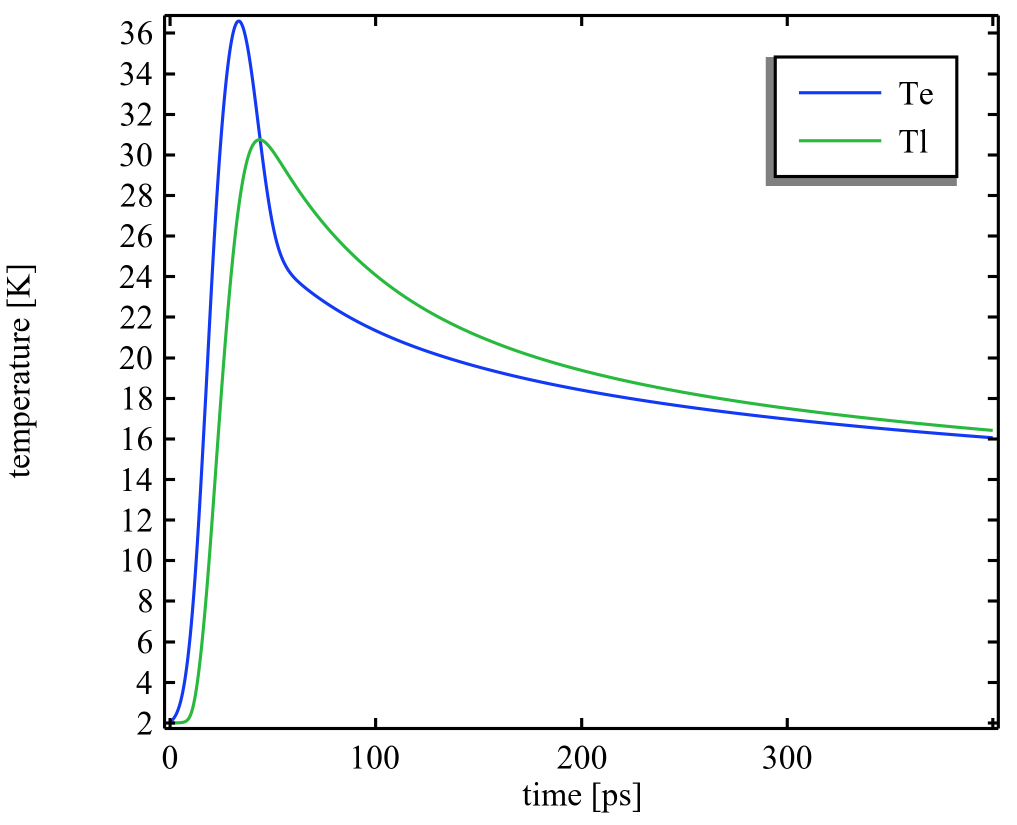}
\caption{Temporal evolution of electron ($T_e$) and lattice ($T_l$) temperatures at the surface of a copper photocathode, initially at $T_0 = \SI{2}{K}$. The depicted thermal response is calculated for an absorbed laser fluence of $F_{\text{abs}} = \SI{2.2}{J/m^2}$ and an RMS laser pulse duration $\sigma_t = \SI{8}{ps}$.
}
\label{fig:2uj_100pc_1mhz}
\end{figure}

First, let us consider the laser source term (Eq.~\eqref{eq:source_term}) with parameters providing a physically representative laser fluence for the 1D-TTM model. We define it based on the operational use case at the future CW European XFEL~\cite{bazyl2021cw}. For a design specification QE of $2.5 \times 10^{-4}$, generating $\SI{100}{pC}$ electron bunches at a $\SI{1}{MHz}$ repetition rate corresponds to an average laser power of $\SI{2}{W}$ (see the right plot in Fig.~\ref{fig:gun_model}). This implies an incident laser pulse energy of $E_{\mathrm{pulse}} = \SI{2}{\micro\joule}$ (since $P_{\text{avg}} = E_{\text{pulse}} \times f_{\text{rep}}$). The laser spot is treated as having a top-hat intensity profile with a radius of $r_{\mathrm{spot}}=\SI{430}{\micro\metre}$, which corresponds to a spot area $A_{\mathrm{spot}} = \pi r_{\mathrm{spot}}^2$. With a material reflectivity of $R=\num{0.36}$ for copper at an excitation wavelength of $\SI{257}{\nano\metre}$ \cite{PhysRevB.6.4370}, the absorbed fluence is $F_{\mathrm{abs}}=(1-R)E_{\mathrm{pulse}}/A_{\mathrm{spot}}$. The pulse has an rms duration of $\sigma_t = \SI{8}{\pico\second}$ and its peak is centered at $t_c = 4\sigma_t = \SI{32}{\pico\second}$. The effective penetration depth $d_p$, which accounts for both optical absorption and non-thermal electron transport, is taken to be $\SI{83}{\nano\metre}$. This value is calculated as the sum of the optical penetration depth ($\approx\SI{13}{\nano\metre}$) and the ballistic transport range ($\approx\SI{70}{\nano\metre}$), an approach proposed for UV excitation of copper in \cite{Maxson2017}. Figure ~\ref{fig:2uj_100pc_1mhz} shows time dependent temperatures of electrons and lattice at x = 0. For the relatively long $8 \text{ ps}$ RMS laser pulse, combined with a low absorbed fluence of $\SI{2.2}{J/m^2}$, the 1D-TTM model predicts a modest $\SI{6}{K}$ difference between the peak electron temperature ($T_e \approx \SI{36.5}{K}$) and lattice temperature ($T_l \approx \SI{30.5}{K}$).

For the considered design parameter set of the CW photoinjector for the European XFEL, the 1D-TTM does not predict a significant temperature growth and the observed thermal non-equilibrium is modest. It should be noted that the assumed QE is an optimistic assumption. Factors such as a larger pulse energy, a smaller laser spot size, and a shorter pulse duration can have a significant impact on the electron lattice thermal behavior. 

\subsection{\label{ttm_disc}Discussion of the results obtained using the 1D-TTM}
The TTM is predicated on the assumption that following photon absorption, the electron subsystem thermalizes internally via electron-electron scattering on a timescale  $\tau_{e-e}$, that is much shorter than the timescale for electron-phonon energy transfer $\tau_{e-ph}$. This condition ($\tau_{e-e} \ll \tau_{e-ph}$) ensures the existence of a well-defined, quasi-equilibrium electron temperature $T_e$. This assumption is known to be challenged under conditions of very low excitation intensity or with ultrashort (femtosecond) laser pulses. In such cases, the electron distribution may remain non-thermal (i.e., non-Fermi-Dirac) for a period comparable to $\tau_{e-ph}$, making the concept of an electronic temperature technically ill-defined and requiring more fundamental models, such as the Boltzmann Transport  (BTE), for a rigorous description~\cite{PhysRevB.65.214303}. Furthermore, at cryogenic range of $T_l$, the applicability of TTM becomes more fragile: under low fluence and femtosecond excitation, experiments on noble metals showed thermal behavior incompatible with TTM because the electron distribution can remain non-thermal on times scales comparable or longer than excitation source~\cite{PhysRevB.51.11433}. For such operational conditions, the TTM model can be extended by introducing an additional non-thermal electron reservoir $U_{nt}$ to account for finite thermalization effects \cite{Carpene2006,Tsibidis2018, Uehlein2022}. 

However, the laser pulse relevant to this work are on the picosecond scale ($\sigma_t=8$ ps; FWHM $\approx$ 19 ps). This pulse duration is significantly longer than the sub-picoseconds timescales typically associated with electron thermalization in noble metals. The laser does not act as an instantaneous impulse, but rather as a continuous heating source over a period comparable to, or longer than, the internal relaxation times. This allows a quasi-equilibrium state to be established and maintained, where the electron subsystem is simultaneously heated by the laser, thermalized by electron-electron scattering and cooled by the lattice. In this regime, the TTM serves as a valid and physically appropriate model for describing the average energy balance and the coupled thermal dynamics of the electron and lattice subsystems on the relevant picosecond timescale.

The purpose of 1D-TTM in this work is twofold. First, it demonstrates that for the operational parameters of the CW photoinjector, the peak surface temperatures remain modest in the cryogenic range and the non-equilibrium is minimal at the nominal laser fluence. Second, we define an interface at a depth $x_{\text{int}}$ within the 1D TTM domain and calculate the net conductive heat flux crossing the plane. The flux represents the total energy per unit area and time that propagates from the highly non-equilibrium surface layer into the more equilibrated bulk. The resulting time-dependent heat flux serves as a heat source for the macroscopic model. The net flux is the sum of the electronic and lattice contributions:

\begin{equation}
    q_{\text{net}}(t) = \left. \left[ -K_e(T_e,T_l)\frac{\partial T_e}{\partial x} - K_l(T_l)\frac{\partial T_l}{\partial x} \right] \right|_{x=x_{\text{int}}}
    \label{eq:ttm_flux}
\end{equation}
where $x_{\text{int}}=\SI{2}{\mu m}$ which is sufficiently deep to be beyond the direct laser absorption and initial ballistic electron transport zones ensuring the flux represents a cohesive, diffusive thermal wave. A sufficiently long simulation time is required to ensure  global energy conservation.

\begin{figure}[htb] 
\includegraphics[width=0.5\linewidth]{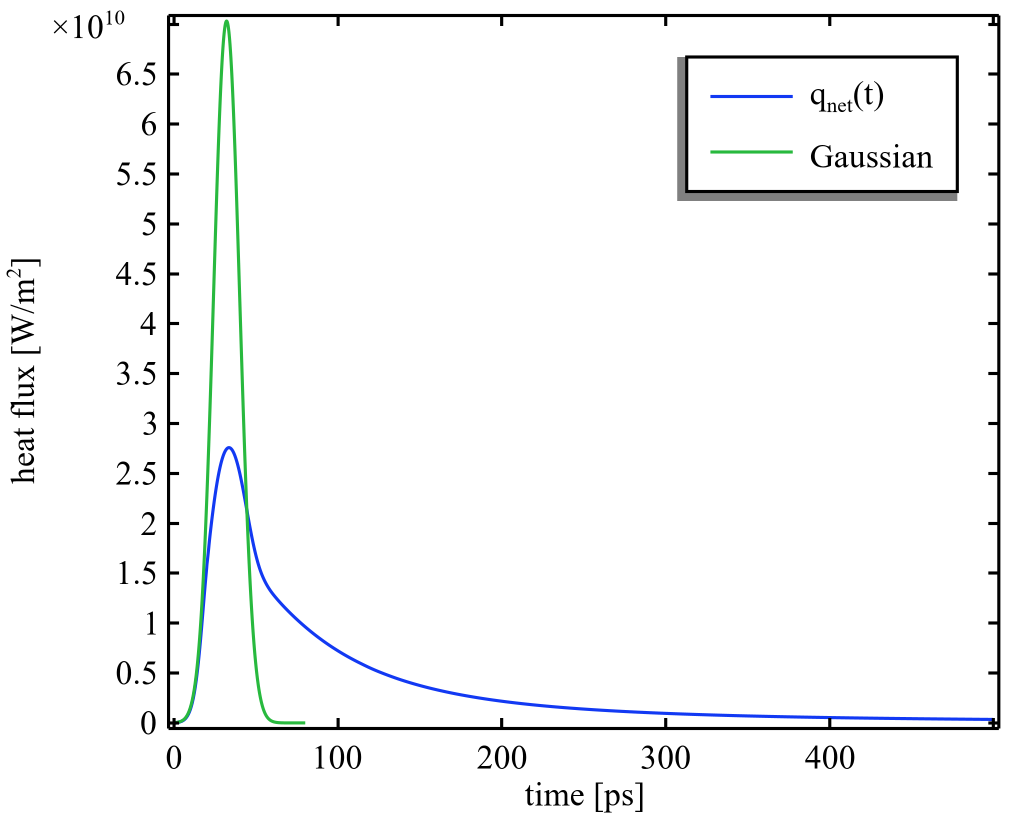} 
\caption{Qualitative comparison of temporal distribution of the heat flux derived from the TTM and a standard 8 ps rms Gaussian laser pulse. Both correspond to a total absorbed pulse energy $\SI{2}{\micro \joule} \times (1-0.36) = \SI{1.28}{\micro\joule}$.}
\label{fig:flux_comp}
\end{figure}

Comparing the temporal distribution of $q_{\text{net}}(t)$ derived from the 1D-TTM to the incident laser's Gaussian profile (Fig. \ref{fig:flux_comp}) reveals that the TTM-based heat source represents a more physically realistic heat deposition, with a more gradual temporal distribution that accounts for the intrinsic energy transfer time within the material.

\section{\label{sec:2dSTM}Two-Dimensional Transient and Stationary Thermal Analysis of a Bulk Copper Photocathode}

\begin{figure}[htb] 
\includegraphics[width=1.0\linewidth]{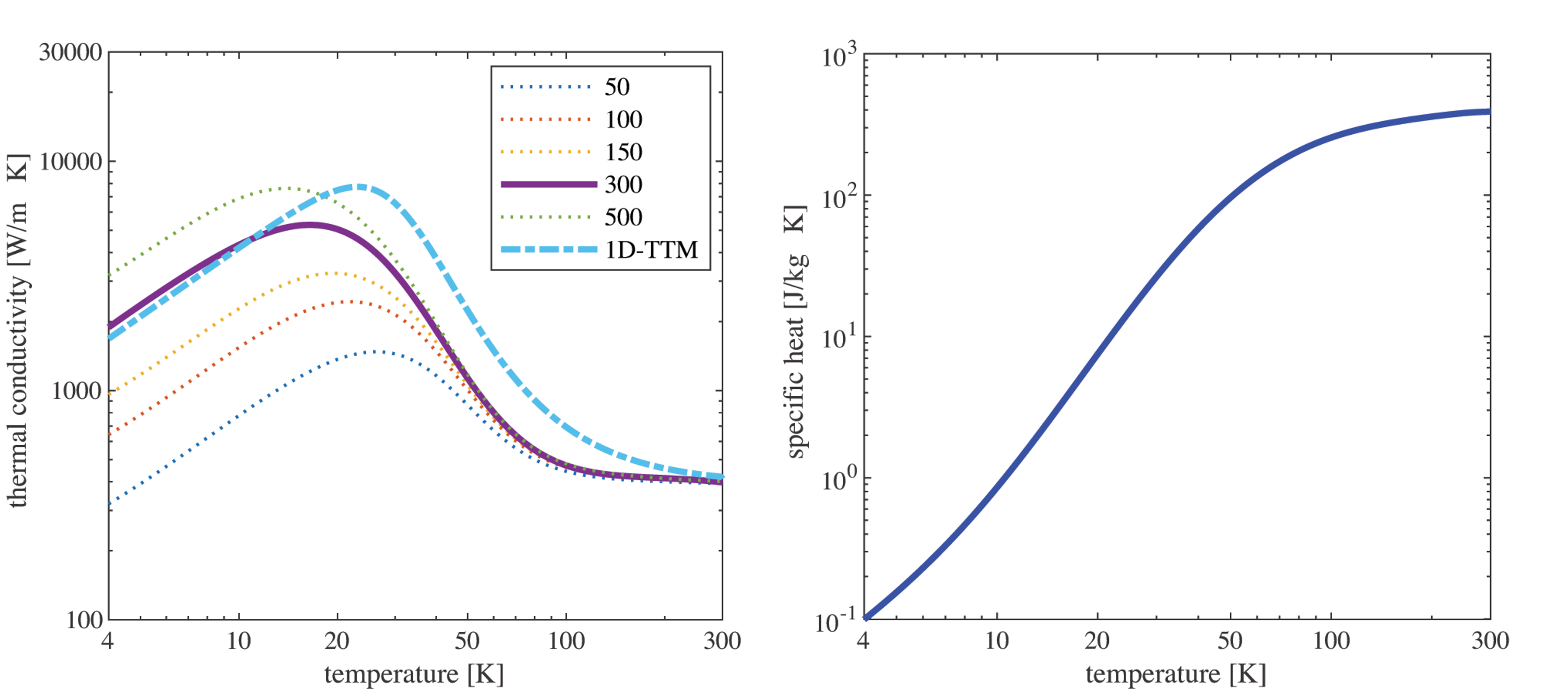} 
\caption{The left plot shows the thermal conductivity of copper as a function of temperature for various values of RRR. The solid line (RRR=300) represents the NIST fit used in the bulk model. The dashed line shows the conductivity derived from the TTM components for comparison. The right plot presents the specific heat capacity of copper as a function of temperature.}
\label{fig:k_Cu}
\end{figure}

We characterize the temperature of the bulk cathode with the aim to validate a steady-state approach for subsequent 3D analysis. For timescales significantly longer than the electron-phonon relaxation time, the thermal behavior within the bulk copper is described by the classical heat conduction equation:
\begin{equation}
\rho(T) C_p(T) \frac{\partial T}{\partial t} = \nabla \cdot (k(T) \nabla T) + Q_{\text{vol}}(r,z,t),
\label{eq:bulk_heat_conduction_transient}
\end{equation}
where $T(r,z,t)$ is the temperature field as a function of the radial coordinate $r$, the axial coordinate $z$, and the time $t$; $\rho$ is the material density; $C_p(T)$ is the temperature-dependent specific heat capacity; $k(T)$ is the temperature-dependent thermal conductivity; and $Q_{\text{vol}}$ is the volumetric heat source. For the bulk model, the temperature-dependent thermal conductivity $k(T)$ and specific heat capacity $C_p(T)$ are implemented using polynomial fits to cryogenic reference data for copper (RRR = 300) from \cite{Simon1992CopperCryo}. The left plot in Fig.~\ref{fig:k_Cu} shows the corresponding thermal conductivity of Cu for various RRR values, with the solid line representing the data used for the bulk cathode modeling. For comparison, we show $K_e$ obtained in Sec.~\ref{sec:mat_prop} assuming $T_e=T_l$. The discrepancy arises because our theoretical model calculates the conductivity from idealized physical laws, while the NIST data are based on a polynomial fit that relies on experimental measurements. The right plot in Fig.~\ref{fig:k_Cu} illustrates the specific heat capacity as a function of temperature for Cu. While the density $\rho$ also exhibits a temperature dependence, its variation for solid-state copper is minimal, and a constant room-temperature value of $\rho = \SI{8960}{\kilogram\per\cubic\metre}$ is assumed.

Solving the time-domain heat transfer model for a $\SI{1}{MHz}$ repetition rate with picosecond laser pulse duration is computationally expensive. In such cases, a stationary  model offers a significantly simplified analysis. The steady-state solution assumes neglecting the time-derivative in Eq.~\eqref{eq:bulk_heat_conduction_transient} and replacing the pulsed heat source with its time-averaged equivalent boundary condition. The primary goal of the subsequent analysis is to validate the steady-state simulation approach, by directly comparing its results to those from a full transient simulation.

\subsection{Validation of the steady-state approach}

\begin{figure}[htb]
\includegraphics[width=1.0\linewidth]{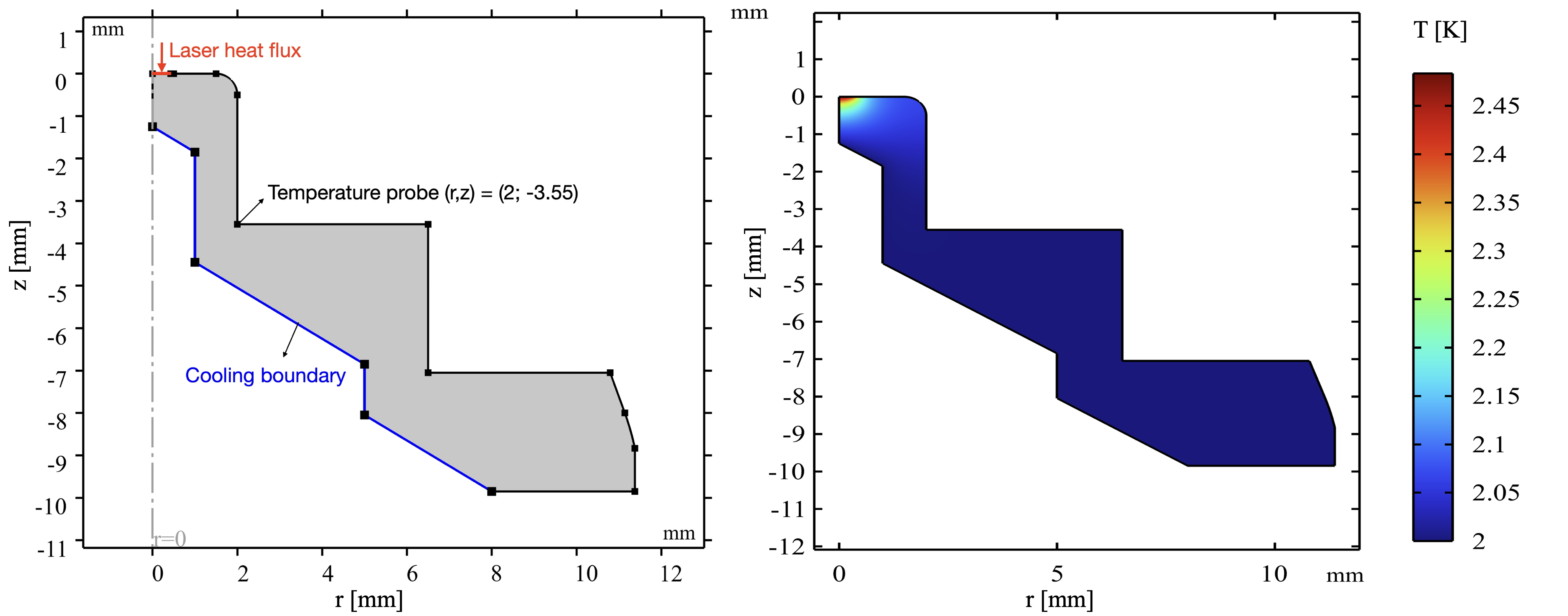}
\caption{The left image shows the simplified axially symmetric representation of the Cu cathode plug with partially indicated boundary conditions: blue cooling boundary, red corresponds to the laser heat flux and the remaining boundary conditions are set to a Neumann boundary condition. The right image presents the steady-state temperature distribution in the model with ideal cooling, for an average incident laser power of $2\,$W.}
\label{fig:ax_sym_mod}
\end{figure}

We validate the steady-state approach by using a simplified two-dimensional, axially symmetric model with the boundary conditions as shown in Fig.~\ref{fig:ax_sym_mod}.  

\begin{figure}[htb]
\includegraphics[width=0.5\linewidth]{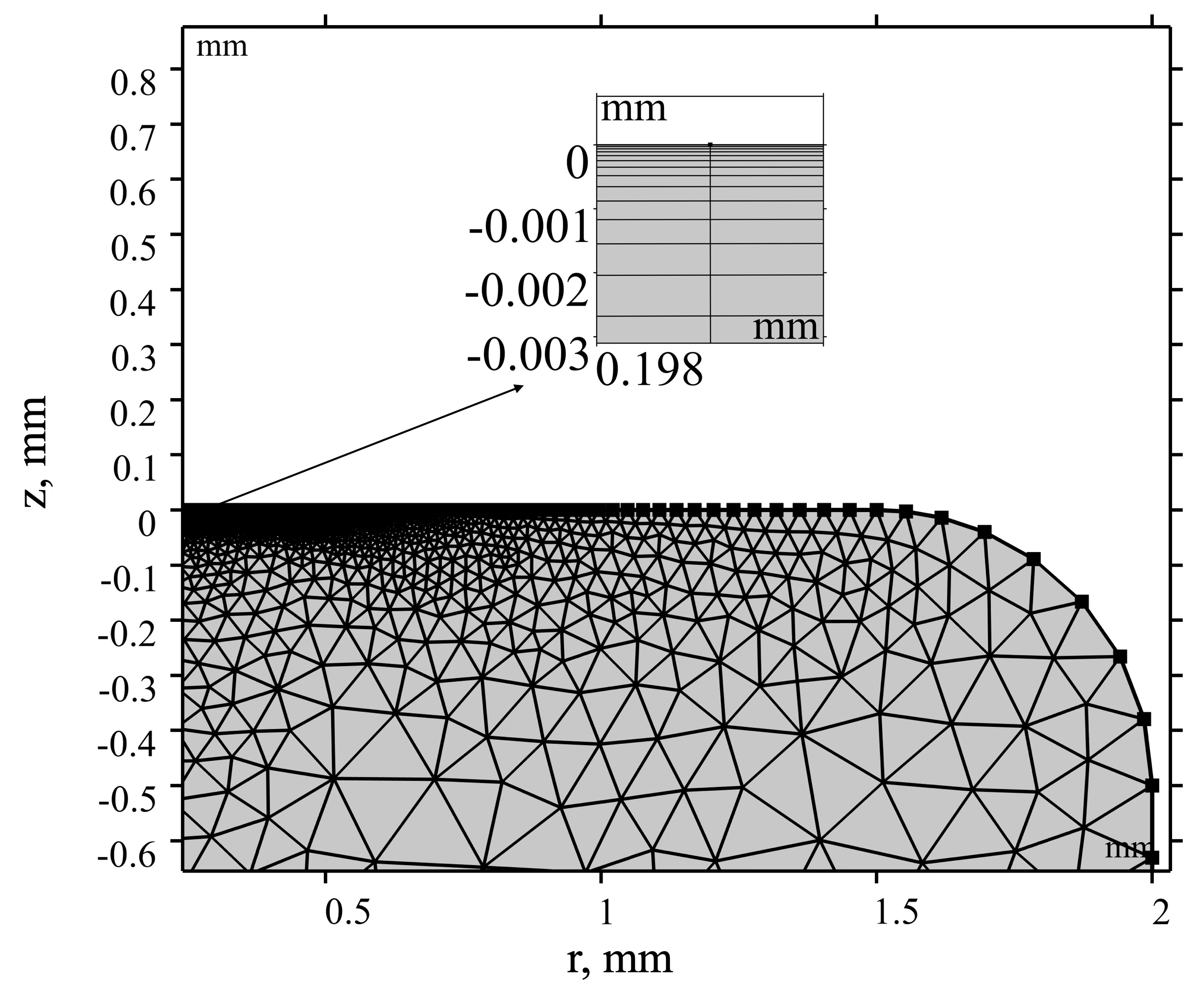}
\caption{Mesh refinement at the boundary layer near the laser heat flux boundary condition.}
\label{fig:mesh}
\end{figure}

 We begin by solving the transient heat conduction problem described by Eq.~\eqref{eq:bulk_heat_conduction_transient}. The heat source from the laser is modeled as a periodic boundary heat flux applied at the cathode emission surface ($z = 0, r \le r_{\text{spot}}=\SI{430}{\micro m}$)
\begin{equation}
    -\mathbf{n} \cdot (k \nabla T) = q_{\text{net}}(t),
    \label{eq:heat_flux_TTM}
\end{equation}
where $q_{\text{net}}(t)$ is the time-dependent heat flux profile obtained from the 1D-TTM analysis, as defined in Eq.~\eqref{eq:ttm_flux}. A primary challenge in the achieving a numerical convergence of the solution in time-domain is resolving steep thermal gradients near the boundary, where the heat flux is applied. To address this, a locally refined mesh is used within the boundary layer region, where the solution varies significantly, as illustrated in Fig. \ref{fig:mesh}.

Next, we solve the corresponding steady-state problem using a time-averaged inward heat flux boundary condition:
\begin{equation}
    -\mathbf{n} \cdot (k \nabla T) = q_{\text{avg}},\quad q_{\text{avg}} = P_{\text{avg}}(1-R)/A_{\text{spot}},
    \label{eq:heat_flux_steady_state}
\end{equation}
where $P_{\text{avg}}$ denotes the incident laser power, and $R = 0.36$ is the material reflectivity. 

\begin{figure}[htb]
\includegraphics[width=0.5\linewidth]{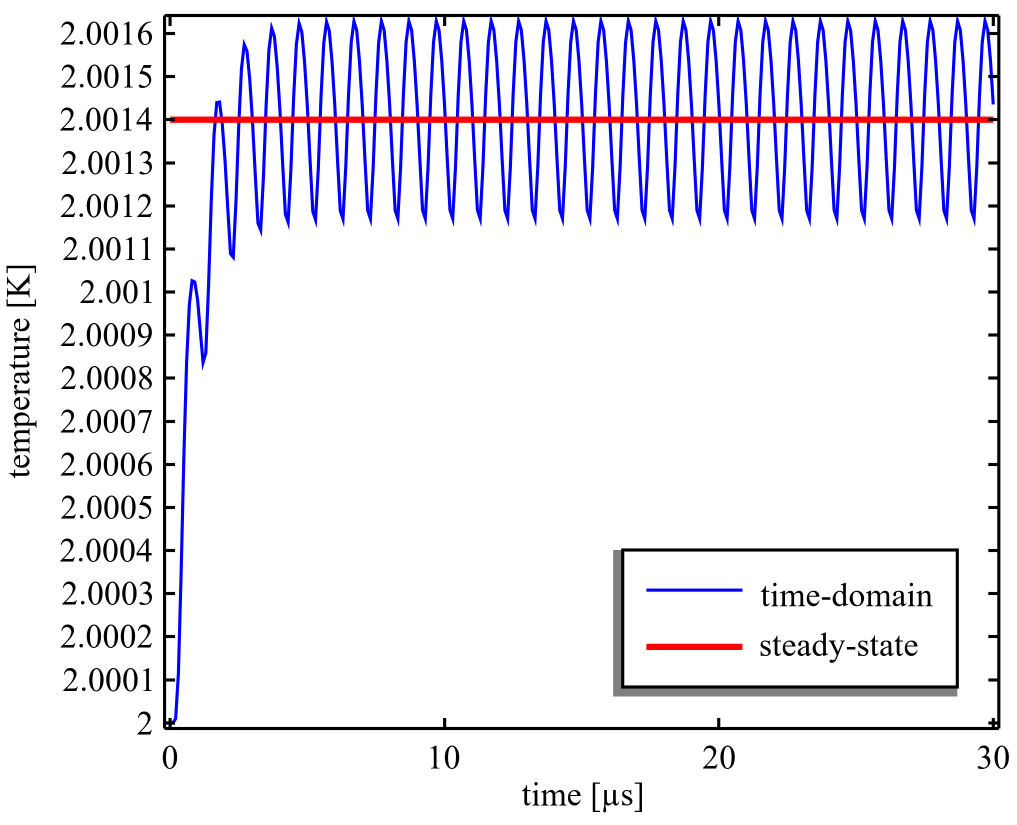}
\caption{Temperature versus time from the transient model, probed at the foreseen initial Cu-In-Nb contact point ($r=2, z=-3.55$) under ideal cooling ($T_0=2\,$K) and periodic $1\,\mathrm{MHz}$ laser heating with $2\,\mu\mathrm{J}$ incident energy per pulse.}
\label{fig:time_domain_2D}
\end{figure}

Fig.~\ref{fig:time_domain_2D} shows the transient solution where the temperature evolution is evaluated at the point of interest (2; -3.55) (see Fig.~\ref{fig:ax_sym_mod}). After a ramp-up period of approximately $\SI{6}{\micro\second}$, the temperature stabilizes into an oscillation with a mean value of $\SI{2.0014}{K}$. The steady-state solution at the same point is $\SI{2.0014}{K}$, thus, matching precisely the transient result. The comparison indicates that the thermal time constant of the bulk cathode is significantly longer than the pulse period, causing the material to respond primarily to the average power input. 

This validation confirms that the computationally efficient steady-state model is sufficient for determining the mean operational temperatures of the bulk cathode. For the initial comparative study, an ideal cooling condition ($T=\SI{2}{K}$) was imposed to ensure numerically fast thermal equilibration in the transient solution. In the subsequent analysis, we use the steady-state approach to study a physically realistic thermal model that incorporates the dominant thermal impedance of the Kapitza boundary resistance at the helium-cooled surfaces~\cite{Swartz1989}.

\section{\label{sec:coupled}Three-dimensional  Coupled Thermal-Electromagnetic\\ Analysis}

A central concern for the integrated cathode design is whether the laser-induced heat load negatively affects the injector cavity's thermal and electromagnetic stability. A three-dimensional (3D) steady-state model is developed to evaluate two primary operational limits of the integrated cathode design under laser heat load. The first is a cryogenic stability at Cu-He interface. The second risk factor arises from a positive thermal feedback loop, where heat conducted from the cathode increases the niobium's temperature-dependent surface resistance $R_s(T)$, thus amplifying the local RF power dissipation, which  directly impacts the cavity's quality factor. On the other hand, the structure of the TM$_{010}$ accelerating electromagnetic mode features a minimal magnetic field responsible for RF losses in the region of niobium which is affected by laser heat deposited from the cathode. To determine which effect dominates, we developed and solved a 3D steady-state coupled thermal-electromagnetic model.

\subsection{Thermal-Electromagnetic Model Definition}
\begin{figure}[htb]
\includegraphics[width=0.7\linewidth]{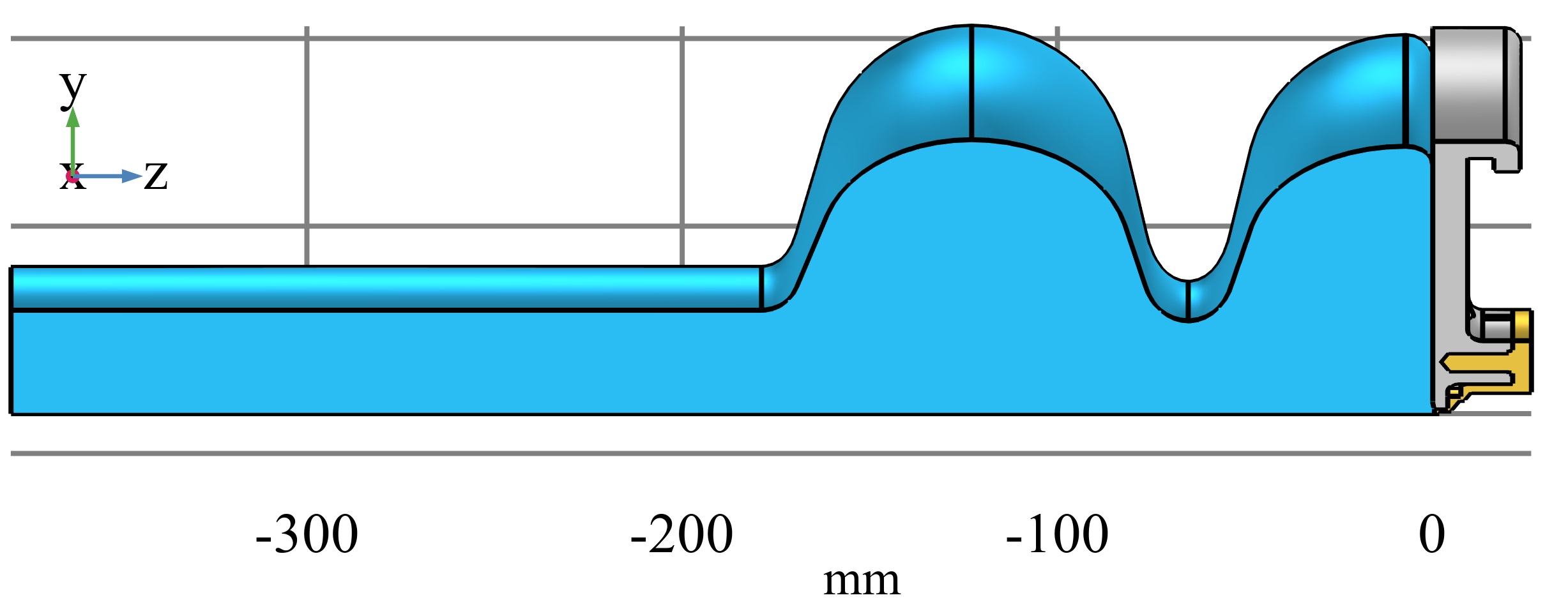}
\caption{1.6-cell SRF injector's 3D model for coupled RF-thermal analysis. The axially symmetric section (1/8) shows the Cu plug (yellow), the Nb back wall (grey), and the RF vacuum volume (blue).}
\label{fig:3D_model}
\end{figure}

\begin{figure}[htb]
\includegraphics[width=1.0\linewidth]{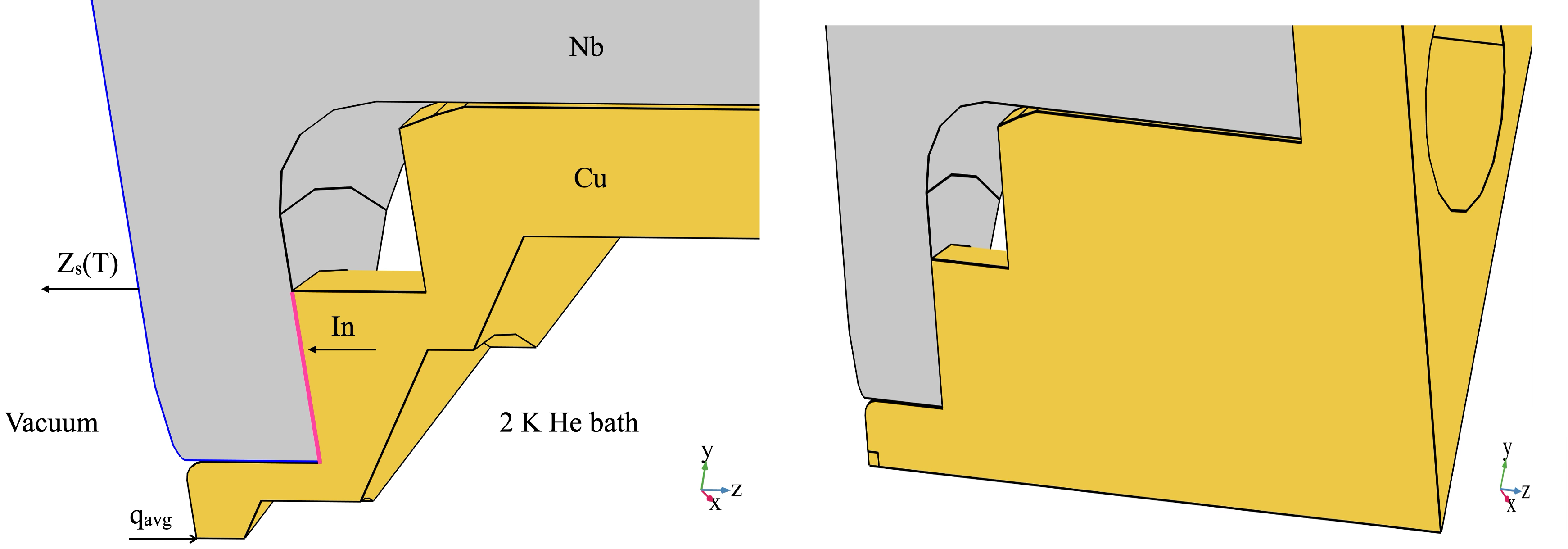}
\caption{The image on the left shows the schematic of the domains and principal boundary conditions in the 3D model at the Cu--Nb interface. The image on the right presents the improved copper cathode plug geometry (yellow).}
\label{fig:3D_model_boundary}
\end{figure}

The computational model, shown in Fig.~\ref{fig:3D_model}, consists of the copper photocathode plug, the niobium cavity back wall, and the RF vacuum volume. A 1/8th section of the geometry is simulated due to the partial axial symmetry of the back wall's mechanical design. A schematic illustrating computational domains and boundary conditions is shown in the left image of Fig.~\ref{fig:3D_model_boundary}.

\begin{figure}[htb]
\includegraphics[width=1.0\linewidth]{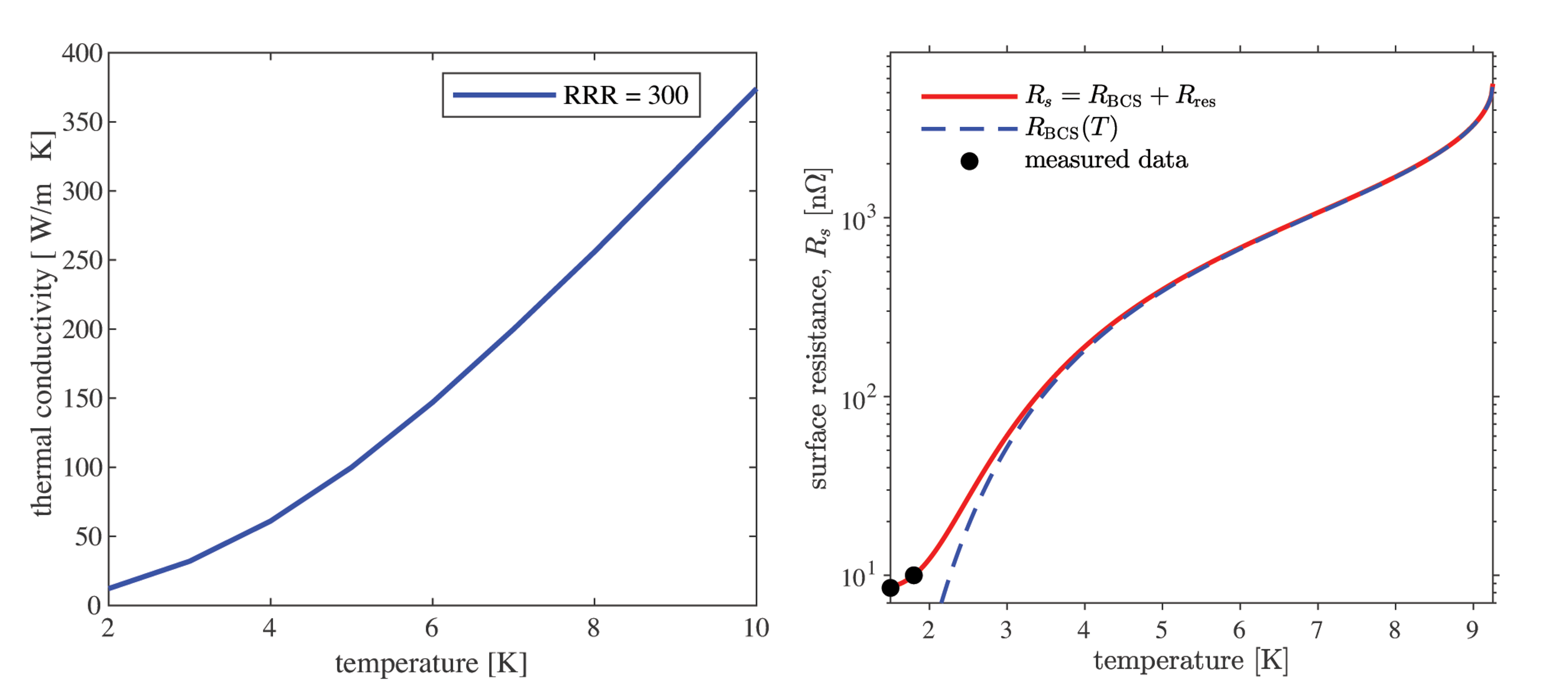}
\caption{The left plot shows the thermal conductivity of niobium (RRR=300) versus temperature \cite{Schilcher1995}. The right plot presents the surface resistance of niobium at $1.3\,$GHz. Points are experimental data for the DESY 16G09 cavity; the red line is the model fit from Eq.~\eqref{eq:bcs_model}.}
\label{fig:k_Nb}
\end{figure}

The thermal analysis is governed by the steady-state heat version of equation Eq.~\eqref{eq:bulk_heat_conduction_transient}. The laser heat is introduced as an inward heat flux boundary condition as defined in Eq.~\eqref{eq:heat_flux_steady_state}, evaluated for an average incident laser power $P_{\text{avg}}$. The temperature-dependent thermal conductivities for copper (RRR=300, Sec.~\ref{sec:2dSTM}) and niobium (RRR=300, shown in the left plot of  Fig.~\ref{fig:k_Nb}) are used. A primary heat transfer path from the cathode to the niobium is the Cu-In-Nb contact, which is characterized by a thermal contact resistance $R_{tc}$. Specific experimental data for the thermal contact resistance of a Cu-In-Nb connection are not readily available. We describe this interface using a value derived from experimental measurements of a similar pressed Nb-In-Al joint for which Dhuley \textit{et al.} measured a contact resistance of $R_{tc} \approx \SI{1.0e-4}{\kelvin\metre\squared\per\watt}$
  at a temperature of $4.5\,$K~\cite{DHULEY201886}. The thermal resistance of such solid-solid contacts, when limited by phonon transport at cryogenic temperatures, typically follows an inverse cubic scaling $R_{tc} \propto T^{-3}$. Extrapolating the measured $4.5\,$K value down to the operational temperature of the injector yields a resistance an order of magnitude higher. Therefore, for this analysis, we adopt a conservative estimate for the area-specific contact resistance of the Cu-In-Nb joint $R_{tc, \text{Cu-In-Nb}}(2\,\text{K}) \approx \SI{1.0e-3}{\kelvin\metre\squared\per\watt}$.
This thermal resistance is applied at the indium-sealed interfaces as a thermal contact resistance boundary condition without directly introducing  indium into the computational domain. This imposes a temperature discontinuity $\Delta T = T_u - T_d$ that is proportional to the normal heat flux $q_n$ flowing across the interface:
\begin{equation}
	q_{n} = \mathbf{n} \cdot (-k \nabla T), \qquad T_u - T_d = q_n \cdot R_{tc}
	\label{eq:Rtc_boundary_condition}
\end{equation}
where $T_u$ and $T_d$ are the temperatures on the upstream and downstream sides of the boundary, respectively. $R_{tc}$ is the area-specific thermal contact resistance and the heat flux $q_n$ is continuous across the boundary.

The cooling regime at the Cu-\text{He\,\textsc{II}} and the Nb-\text{He\,\textsc{II}} is governed by the Kapitza conductance $h_{K}=aT^{3}$, a phenomenon of phonon transport across a boundary. The standard physical model for this process across a temperature difference is provided by the modified acoustic mismatch theory~\cite{Swartz1989}. Table~\ref{tab:kapitza_coefficients} summarizes typical experimental values for the pre-factor $a$ [W\,m$^{-2}$\,K$^{-4}$]. For the copper photocathode in our model, we assume a clean surface while for the niobium cavity, we use a value representative of a standard buffered chemical polish (BCP) finish. It is often convenient to describe this boundary impedance in terms of the Kapitza thermal resistance $R_K(T) = [h_K(T)]^{-1}$. The temperature-dependent Kapitza resistance is implemented as the cooling boundary condition for the steady-state thermal analysis:

 \begin{equation}
     q = -\mathbf{n} \cdot (k \nabla T) = \frac{T - T_{\text{bath}}}{R_K},
    \label{eq:kapitza_bc}
 \end{equation}
 where $\mathbf{n}$ is the normal vector, $T$ is the local surface temperature, and $T_{\text{bath}} = \SI{2}{K}$ is the helium bath temperature.
\begin{table}[htb]
\caption{Kapitza conductance coefficient, $a$, for Cu--\text{He\,\textsc{II}} and Nb--\text{He\,\textsc{II}} where the heat conductance is 
$h_{K}=aT^{3}$ at $T \approx 2\,$K.  Values are representative mid‑points compiled from the literature.}

\label{tab:kapitza_coefficients}
\begin{ruledtabular}
\begin{tabular}{llc}
\textbf{Material} & \textbf{Surface Condition} & \textbf{$a$ (W\,m$^{-2}$\,K$^{-4}$)} \\
\hline
Copper & Clean                 & $1.0 \times 10^{3\,\textsuperscript{a}}$ \\
       & Technical             & $5.0 \times 10^{2\,\textsuperscript{b}}$ \\
       & Rough                 & $2.0 \times 10^{2\,\textsuperscript{a}}$ \\[4pt]
Niobium& Electropolished       & $1.0 \times 10^{3\,\textsuperscript{c}}$ \\
       & BCP cavity‑grade      & $4.0 \times 10^{2\,\textsuperscript{d}}$ \\
       & Rough                 & $2.0 \times 10^{2\,\textsuperscript{d}}$ \\
\end{tabular}
\end{ruledtabular}
\begin{flushleft}
\footnotesize
\textsuperscript{a}Lebrun and Tavian\,\cite{lebrun2014cooling};\;
\textsuperscript{b}Swartz and Pohl\,\cite{Swartz1989};\;
\textsuperscript{c}Amrit \emph{et al.}\,\cite{amrit2000thermal};\;
\textsuperscript{d}Aizaz \emph{et al.}\,\cite{aizaz2010phonon} and Dhakal \emph{et al.} \cite{PhysRevAccelBeams.20.032003};\;

\end{flushleft}
\end{table}

The computed steady-state temperature distribution provides the input for the electromagnetic eigenmode analysis. The coupling to the electromagnetic computation domain is implemented through a surface impedance boundary condition with impedance $Z_s(T) = R_s(T) + i X_s(T)$ applied at the backwall of the cavity. The remaining boundaries for the EM problem are treated as perfect electric conductors (PEC). The surface reactance $X_s$ is calculated for niobium at a frequency $f = \SI{1.3}{\giga\hertz}$ using the London penetration depth approximation $X_s \approx \omega \mu_0 \lambda_L\approx\SI{0.40}{m\ohm}$, where the angular frequency is $\omega = 2\pi f$, $\mu_0$ is the permeability of free space, and the London penetration depth is taken as $\lambda_L \approx \SI{39}{\nano\meter}$. While temperature dependent, $X_s$ is approximated as a constant value due to the localized domain affected by cathode laser heating. The resistive component $R_s(T)$ is the sum of a residual component $R_{\text{res}}$ and the BCS resistance, given by:
\begin{equation} \label{eq:bcs_model}
    R_{BCS}(T) = \frac{C}{T} f^2 \exp\left(-\frac{\Delta(T)}{k_B T}\right).
\end{equation}
The superconducting energy gap, $\Delta(T)$, is computed using the Muhlschlegel approximation with $T_c = 9.25\,$K and $\Delta(0)/k_B T_c = 1.9$. The free parameters, $C$ and $R_{\text{res}}$, were determined via a direct fit to measurements of the CW SRF injector cavity at DESY (16G09) (the right plot of Fig.~\ref{fig:k_Nb}), yielding $R_{\text{res}} = 8.3\,\text{n}\Omega$ and $C = 31182\,\text{n}\Omega\cdot\text{K}/\text{GHz}^2$.

\subsection{Thermal Stability at the Copper-Helium Interface}
\begin{figure}[htb]
    \centering
    \includegraphics[width=1.0\linewidth]{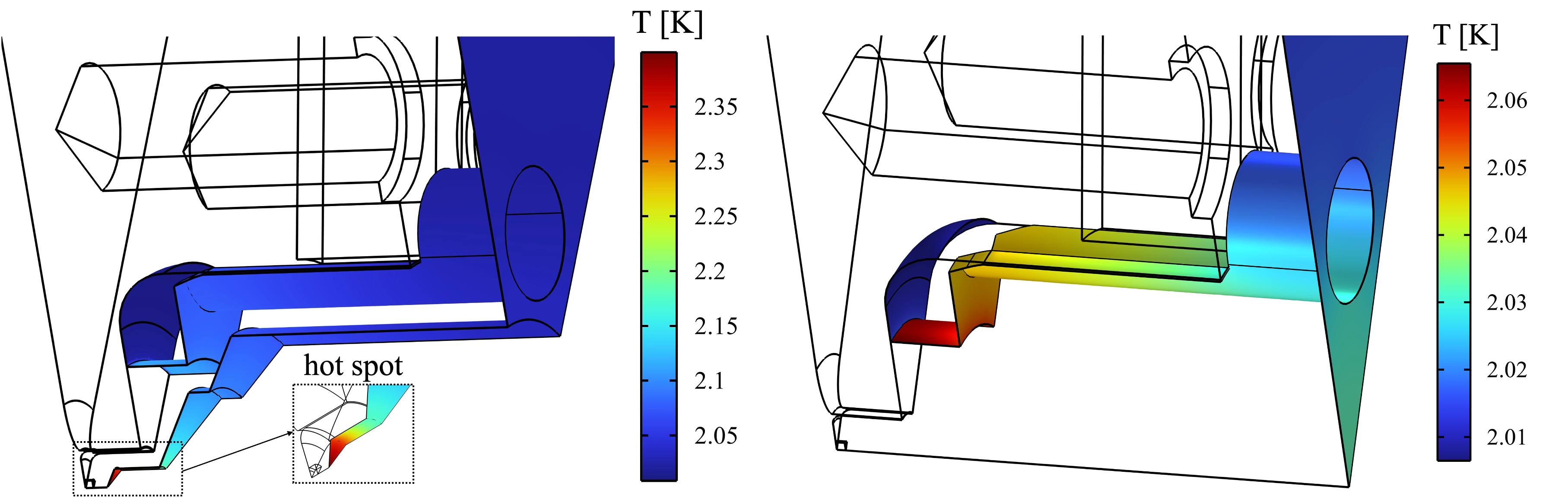}
    \caption{The image on the left shows the steady-state temperature distribution on the helium-wetted surfaces of the copper plug and adjacent niobium back wall, for an incident laser power of $2\,$W. The inset highlights the localized hot spot on the geometric protrusion. The image on the right presents the steady-state temperature distribution on the helium-wetted surfaces of the improved version of the copper plug and adjacent niobium back wall, for an incident laser power of $2\,$W.}
    \label{fig:hot_spot_T}
\end{figure}

The analysis of the 3D model with the baseline cathode geometry shown in the left image of Fig.~\ref{fig:3D_model_boundary} reveals a potential cooling issue. For a nominal incident laser power of 2 W, the steady-state solution predicts that the temperature on the helium-wetted copper surface exceeds the helium saturation temperature ($T_\text{sat} \approx 2.0$~K at $\sim 31$~mbar) by $\Delta T = T_\text{peak} - T_\text{sat} \approx 0.4$~K.  The left image in Fig.~\ref{fig:hot_spot_T} illustrates this for an incident laser power of $2\,$W, indicating a localized hot spot on a conical protrusion reaching approximately $2.4\,$K.  Here $T_\text{peak} \approx 2.4$~K denotes the copper temperature, while the helium side is numerically fixed at $T_\text{bath} = T_\text{sat} = 2$~K by the Kapitza-to-bath boundary in Eq.~\ref{eq:kapitza_bc}. This boundary condition is only valid for single-phase He-II. However, helium boiling can arise when the helium-side peak heat flux limit is approached or exceeded, with the threshold depending on geometry and immersion depth \cite{vanSciver2012helium, PhysRevB.108.174509}.  Within this modeling framework, the magnitude of $\Delta T$ is used as a conservative indicator of a thermally loaded interface where two-phase effects could potentially develop at the indicated hot spot. The present model indicates risk but does not predict onset or the critical heat flux (CHF). 

The potential onset of two-phase effects fundamentally alters the heat transfer mechanism from efficient solid-to-superfluid conduction to less predictable two-phase regime. The primary concern in this regime is the onset of film boiling, which insulates the hot spot with a layer of helium vapor. If the insulating vapor film propagates, heat transport into the helium bath can be strongly reduced and a larger fraction of the load conducts into the SRF injector's niobium structure. Furthermore, sustained boiling dynamics near a high-Q SRF cavity introduces a secondary parasitic effect, i.e. it leads to thermomechanical vibrations~\cite{McGee2016}.

The thermal performance of the baseline plug geometry is constrained by a conductive bottleneck within the plug geometry. While cooling at the boundary is initially governed by Kapitza resistance, the narrow cross-section of the protrusion creates a significant internal thermal impedance. This impedance concentrates the heat flux, leading to the localized temperature rise before heat can be effectively distributed across the wetted interface. An alternative cathode plug geometry, shown conceptually in the right image of Fig.~\ref{fig:3D_model_boundary}, was evaluated to address this conductive limitation. The modification increases the local cross-section in the vicinity of the hot spot, providing a lower impedance path for lateral heat conduction that more effectively distributes the thermal load.

The 3D steady-state model was solved for the improved geometry. The right image in Fig.~\ref{fig:hot_spot_T} shows the resulting temperature distribution at the cooling surfaces for a nominal incident laser power of $P_{\text{avg}}=\SI{2}{W}$. For this case, the peak temperature of the copper cathode at the helium interface is $T_{\text{peak}} \approx   \SI{2.07}{K}$, and the corresponding heat flux is approximately $\SI{550}{W/m^2}$ (single-phase Kapitza boundary flux). The $\SI{70}{mK}$ temperature rise is modest and the wetted-copper surface temperature remains near saturation temperature. We then evaluated a higher heat-load case with an average incident laser power of $P_{\text{avg}}=\SI{5}{W}$. The simulation predicts a peak helium-wetted surface temperature of $T_{\text{peak}} \approx \SI{2.14}{K}$. For the subsequent RF analysis, we treat the 5 W temperature field as single-phase cooling and assume that the Kapitza boundary condition remains valid.

\subsection{Laser heating and RF performance}

We now evaluate the impact of the laser induced temperature gradients on the RF performance of the injector cavity with improved geometry of the Cu cathode plug (see Fig. \ref{fig:3D_model_boundary}, right). The analysis solves the stationary, coupled thermal-electromagnetic model to determine, if the localized heating of the niobium back wall leads to an increase in dissipated RF power.

\begin{figure}[htb]
\includegraphics[width=1.0\linewidth]{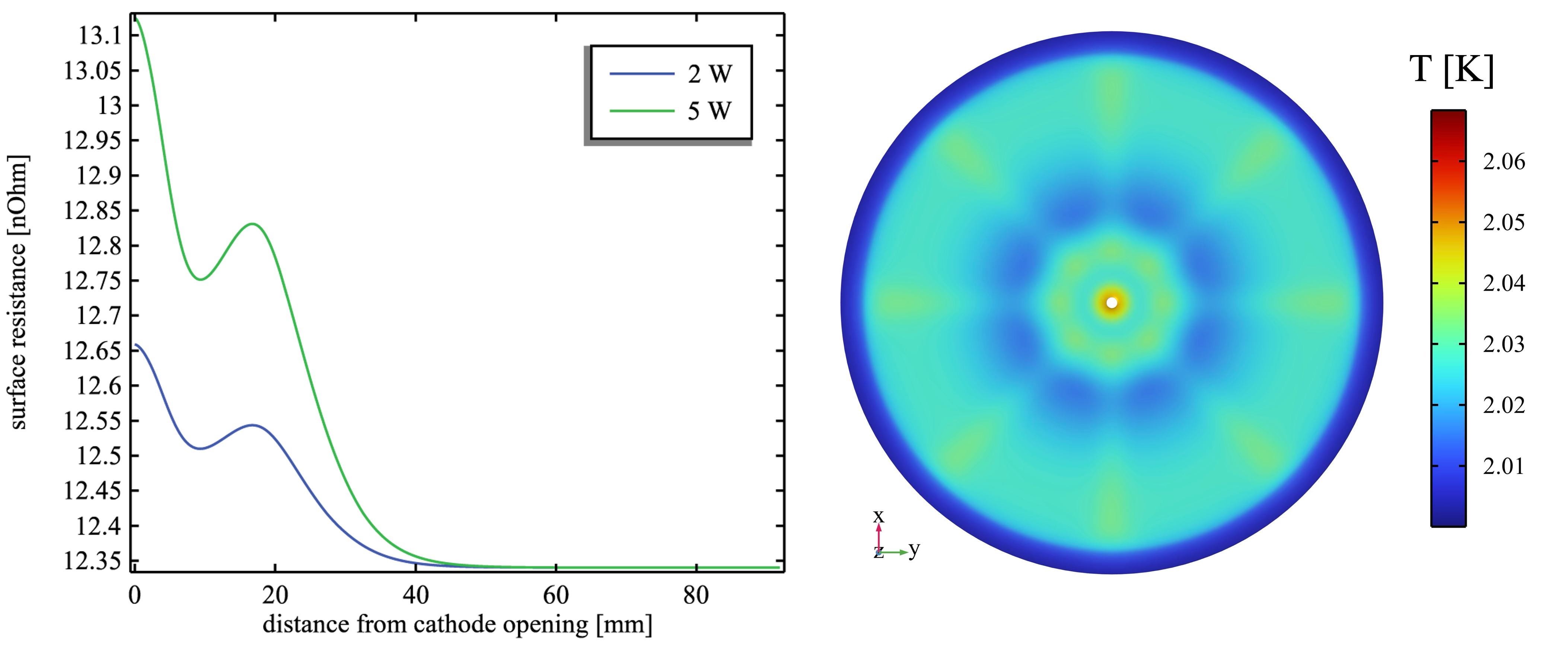}
\caption{The left plot shows the local surface resistance along the radius of the niobium back wall, based on the temperature profiles from the thermal model. The image on the right presents the steady-state temperature distribution on the interior surface of the niobium back wall of the injector cavity, based on the solution of the 4th iteration of the coupled RF-thermal model for an incident laser power of $P_{\text{avg}} = \SI{5}{W}$. }
\label{fig:3d_res_calc}
\end{figure}

\begin{figure}[!htb]
\includegraphics[width=1.0\linewidth]{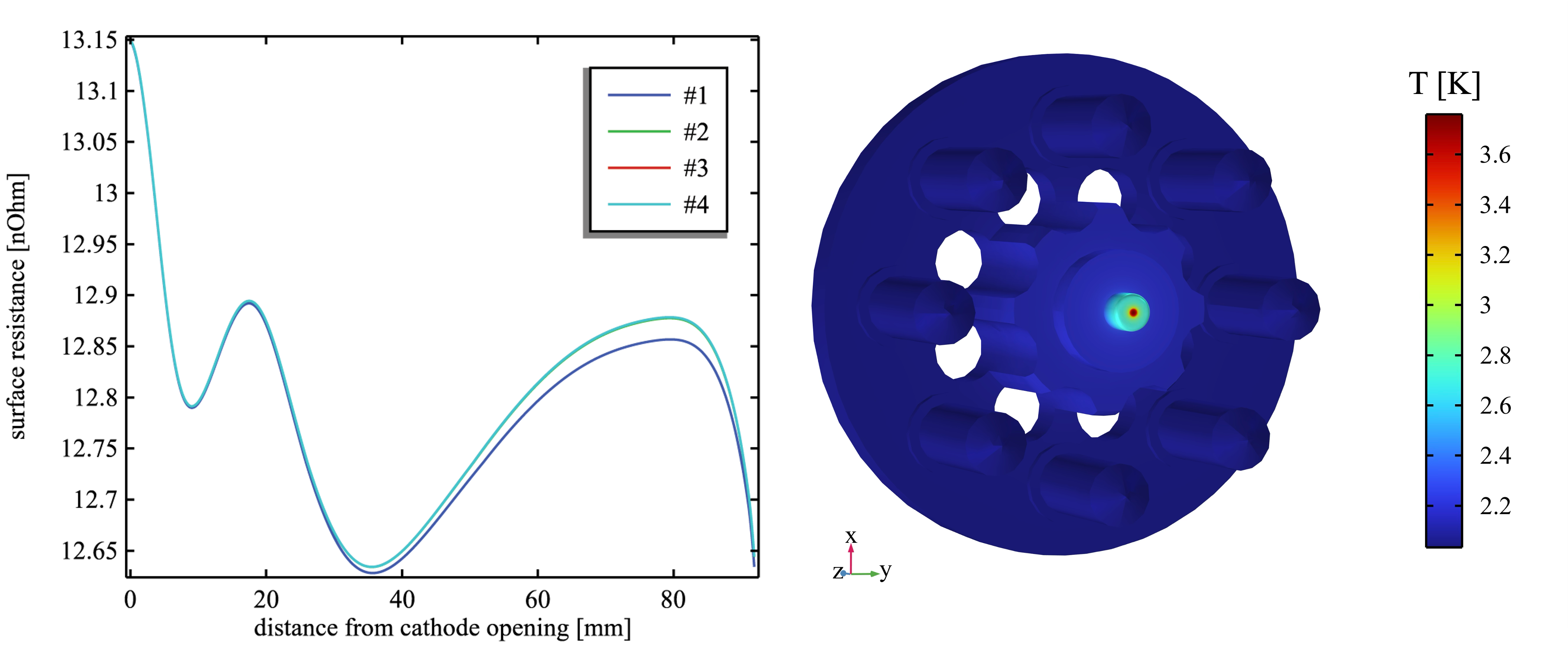}
\caption{The left plot shows the surface resistance profiles along the niobium back wall for four consecutive iterations of the coupled thermal-electromagnetic model, calculated for an incident laser power of $5\,$W. The image on the right presents the steady-state temperature distribution in the copper photocathode plug, based on solution of the 4th iteration of coupled RF-thermal model for an incident laser power of $P_{\text{avg}} = \SI{5}{W}$.}
\label{fig:iterations_resistance}
\end{figure}

First, the thermal model is solved for incident laser powers of $2\,$W and $5\,$W. The temperature field is then used to compute the local surface resistance, shown on the left in Fig.~\ref{fig:3d_res_calc}, where a modest increase is observed near the central axis. Despite the moderate laser induced thermal load of the injector cavity, given the exponential dependence of the BCS resistance on temperature, a multi-step iterative analysis was performed to ensure that the initial temperature rise does not trigger a thermal feedback runaway. The calculated RF losses were added back into the thermal model as a heat source, and the loop was repeated 4 times. The results, shown in Fig.~\ref{fig:iterations_resistance}, demonstrate immediate convergence after the second iteration, confirming the thermal feedback is negligible and the system is electromagnetically stable. The image on the right in Fig.~\ref{fig:iterations_resistance} presents the steady-state temperature distribution in the copper photocathode plug, based on the solution of the 4th iteration of the coupled RF-thermal model for an incident laser power of $P_{\text{avg}} = \SI{5}{W}$.

The temperature distribution in the system is dictated by the copper cathode plug, resulting in a modest temperature rise in the vicinity of the cathode opening of the injector cavity. Moreover, the magnetic field of the TM$_{010}$ $\pi$ mode on the interior of the injector's backwall field is zero on the central axis ($r=0$) and increases with radius. A comparison with the temperature map (the image on the right in Fig.~\ref{fig:3d_res_calc}), indicates that the area of the maximum temperature increase coincides with the region of low magnetic field. As a result, the integral for total dissipated RF power (Eq.~\eqref{eq:pd_q0}) is dominated by the vast surface areas operating at nominal resistance, rendering the effect of localized heating negligible. Calculating the dissipated power assuming first T = 2 K = const. yields 1.109 W at the backwall of the cavity (0th iteration). Following the 4th iteration of thermal-electromagnetic coupled simulation yields a 2.8 \% higher dissipated power of 1.14 W, which is mainly driven by the RF load (50 MV/m axial peak electric field) following the temperature rise after the first iteration. Therefore, we conclude that the primary thermal consideration for the injector cavity design is the management of the cryogenic cooling at the Cu-He interface, rather than the RF-thermal feedback.

\section{Conclusion}
In this work, we developed and applied a comprehensive numerical framework to analyze the thermal characteristics of a cryogenically cooled copper photocathode directly integrated into an SRF injector cavity. The ultrafast surface thermal dynamics were characterized using a one-dimensional two-temperature model developed for the cryogenic regime. For the analysis of bulk heat transfer, a steady-state, time-averaged thermal model was validated against full transient simulations, confirming its computational efficiency. Using this steady-state model, a coupled three-dimensional thermal-electromagnetic analysis was performed, incorporating experimentally fitted data for the niobium surface resistance and literature-based estimates for Kapitza conductance at Nb--He\,\textsc{II} and Cu--He\,\textsc{II} interfaces.

The analysis identified a potential cryogenic instability in the baseline cathode geometry, where even at the nominal laser power ($2\,$W) it is predicted to heat the helium-wetted Cu interface above the helium saturation temperature by about $0.4\,$K at the operating bath pressure ($\sim 31$ mbar). The improved design was shown to maintain the interface temperature near saturation ($T_{\mathrm{peak}}\approx2.07$~K $\Delta T\approx0.07$~K) for the nominal laser power. Operation beyond the nominal load requires dedicated cryogenic analysis (helium-side peak heat flux) and further optimization of the plug geometry to preserve margin against two-phase effects. The precise onset of boiling in He-\textsc{II} depends on the helium-side peak heat flux, which is not predicted by the present model. We have demonstrated that the present design of the cathode plug can benefit from further systematic cryogenic design optimization in order to mitigate potential instabilities related to localized overheating of the helium-wetted copper surface.

Assuming the cryogenic stability of the Cu photocathode is established, we confirm that the SRF injector remains electromagnetically stable. A multi-iteration analysis of the thermal–RF feedback loop at $5\,$W shows rapid convergence. This stability is a direct consequence of the TM$_{010}$ mode's magnetic-field zero on the central axis, which minimizes the impact of localized heating on RF power dissipation in the vicinity of the cathode opening. This finding provides an important design validation and indicates that the primary thermal consideration is the management of the direct cryogenic cooling at the cathode rather than RF–thermal feedback effects. While the modified geometry resolves the cryogenic concern at $2\,$W, stable operation at $5\,$W is dependent on single-phase cooling at the Cu--He interface. Any practical implementation targeting more than $2\,$W should prioritize geometry that reduces spreading resistance and increases the effective wetted perimeter, together with surface conditions that maximize the effective Kapitza conductance, in order to retain an adequate thermal margin.

\section*{ACKNOWLEDGMENTS}

We thank D. Reschke for providing experimental data on surface resistance and thermal conductivity for the injector cavity, as well as for reading the manuscript and providing valuable comments. We thank A.K. Dhillon for providing critical comments on the cryogenic operating conditions of the SRF injector and the associated helium-II heat-transfer physics. We thank M. Wenskat for fruitful discussions on questions of the cryogenic stability in SRF environment. Authors thank W. Decking and H. Weise for their interest and support of this work. We thank H. Sinn, M. S. Tavakkoly and F. Yang (European XFEL GmbH) for useful discussion on ultra-fast laser heating. We thank support team of Comsol Multiphysics GmbH for assistance on technical questions.

\nocite{*}

\bibliography{apssamp}% Produces the bibliography via BibTeX.

\end{document}